# Optimizing In Vivo Data Acquisition for Robust Clinical Microvascular Imaging Using Ultrasound Localization Microscopy

Chengwu Huang[1], U-Wai Lok[1], Jingke Zhang[1], Xiang Yang Zhu[2], James D. Krier[2], Amy Stern[1], Kate M. Knoll[1], Kendra E. Petersen[1], Kathryn A. Robinson[1], Gina K. Hesley[1], Andrew J. Bentall[2], Thomas D. Atwell[1], Andrew D. Rule[2], Lilach O. Lerman[2], Shigao Chen[1]

[1]Department of Radiology, Mayo Clinic College of Medicine and Science, Rochester, MN, USA

[2]Division of Nephrology and Hypertension, Mayo Clinic, Rochester, MN, USA

Corresponding Author: Shigao Chen, Department of Radiology, Mayo Clinic, 200 First Street SW, Rochester, MN 55905, USA Chen.Shigao@mayo.edu

**Funding**: The study was partially supported by the National Institute of Diabetes and Digestive and Kidney Diseases under award numbers of R01DK129205 and R01DK138998, and National Eye Institute under award number of R01EY035084. The content is solely the responsibility of the authors and does not necessarily represent the official views of the National Institutes of Health.

**Competing Interests**: The Mayo Clinic and some of the authors (C.H., JZ and S.C.) have a potential financial interest related to the technology referenced in the research.

ABSTRACT

Ultrasound localization microscopy (ULM) enables microvascular imaging at spatial resolutions beyond the acoustic diffraction limit, offering significant clinical potentials. However, ULM performance relies heavily on microbubble (MB) signal sparsity, the number of detected MBs, and signal-to-noise ratio (SNR), all of which vary in clinical scenarios involving bolus MB injections. These sources of variations underscore the need to optimize MB dosage, data acquisition timing, and imaging settings in order to standardize and optimize ULM of microvasculature. This pilot study investigated temporal changes in MB signals during bolus injections in both pig and human models to optimize data acquisition for clinical ULM. Quantitative indices were developed to evaluate MB signal quality, guiding selection of acquisition timing that balances the MB localization quality and adequate MB counts. The effects of transmitted voltage and dosage were also explored. In the pig model, a relatively short window (approximately 10 seconds) for optimal acquisition was identified during the rapid wash-out phase, highlighting the need for real-time MB signal monitoring during data acquisition. The slower wash-out phase in humans allowed for a more flexible imaging window of 1–2 minutes, while trade-offs were observed between localization quality and MB density (or acquisition length) at different wash-out phase timings. Guided by these findings, robust ULM imaging was achieved in both pig and human kidneys using a short period of data acquisition, demonstrating its feasibility in clinical practice. This study provides insights into optimizing data acquisition for consistent and reproducible ULM, paving the way for its standardization and broader clinical applications.

INTRODUCTION

Ultrasound imaging is a first-line clinical modality for a wide range of applications due to its real-time, noninvasive nature and wide accessibility. However, its resolution is inherently limited by ultrasound diffraction, which is associated with the wavelength of the sound waves. Over the past decade, ultrasound localization microscopy (ULM) has emerged as a novel modality to surpass these resolution limits for in vivo imaging of microvasculature [1-31]. ULM relies on detecting spatially isolated microbubble (MB) contrast agents administered into the bloodstream, localizing them with micrometric precision (up to ten times the resolution of conventional ultrasound), and accumulating thousands of MB positions to reconstruct microvascular images at super-resolved resolution, while maintaining the penetration depth of conventional ultrasound imaging. Beyond mapping vascular morphology, ULM can track MB movement over time to measure flow velocity in 2D or 3D for imaging of microvessel hemodynamics. With significant potential for deep tissue microvascular imaging in vivo, ULM has been extensively explored in both preclinical and clinical investigations [3-6, 13, 14, 26, 32-44]. For detailed reviews of the developments and applications of ULM techniques, refer to the review papers by Couture et al., Christensen-Jeffries et al, Song et al and Dencks et al [32, 33, 45, 46].

Despite its potential, ULM faces inherent limitations in clinical implementation. ULM requires low MB concentration in the bloodstream to enable MB isolation and precise localization in each ultrasound frame [32, 33]. This requirement for low MB concentration prolongs data acquisition times to accumulate enough MB events to fully delineate the microvessel structure. In preclinical studies, this data acquisition process can take tens of seconds or even minutes, facilitated by well-controlled MB concentration with intravenous infusion and minimized tissue motion through anesthesia and mechanically fixed probes. However, clinical imaging conditions are often far from ideal, necessitating shorter acquisition times for most clinical applications using hand-held ultrasound probes [45]. Clinical ULM data acquisition typically needs to be completed within a single breath-hold (preferably less than 10 seconds) to mitigate tissue motion and dataset misalignment from prolonged or multiple acquisitions. While higher MB concentrations can accelerate the filling of the vascular bed with MB, they increase the spatial overlap of MB signals, which is detrimental to localization and tracking for successful ULM [47]. Methods are actively being developed to alleviate this challenge, such as MB separation techniques, deep learning algorithms, or non-localization-based methods to improve imaging performance at high MB concentration [5, 10, 12, 18, 29, 48-57]. Nevertheless, an optimal MB concentration tailored for specific algorithms and imaging settings (such as transmitted frequency, wavelength, and beamforming methods, which are closely associated with the size of the point spread function) is still required for robust microvessel imaging. Furthermore, although continuous infusion is an option, a bolus injection is more practical and commonly used in clinical radiology practice [58, 59]. MB concentration in bolus injection is expected to follow the temporal pattern of time-intensity curve in contrast-enhanced ultrasound (CEUS), and thus dynamically changes over the time course of wash-in and wash-out phases of contrast. Since ULM relies heavily on MB data sparsity, the dosage of bolus MB injection may not necessarily be the same as those recommended for CEUS, and the timing after the MB administration is critical for data acquisition. The dosage and timing of imaging are therefore crucial for the successful, robust, and repeatable implementation of ULM. In addition, an appropriate acoustic output or mechanical index (MI) must be determined to ensure sufficiently high signal-to-noise ratio (SNR) without significantly destructing MBs [58, 59]. The SNR is further related to tissue attenuation, imaging depth, tissue heterogeneity, reverberation, phase aberration, and imaging parameters (frequency, pulse length etc.) associated with image quality, which can vary significantly from case to case in in vivo scanning. The interplay of dosage, acquisition timing, acoustic output, along with the inherent variability in in vivo scanning conditions, collectively makes achieving robust and reproducible ULM challenging in clinical practice.

In this study, we aim to facilitate the optimal clinical implementation of ULM by investigating the changes in MB signals over the course of bolus MB injection in both pig model and human in vivo. We use quantitative indices to assess MB signal quality for MB localization during the entire perfusion course after MB injection to study the inherent trade-off between MB concentration and MB localization quality, to guide optimal data acquisition timing and ULM performance. With continuous imaging of the region of interest (ROI), changes in MB signal quality for localization over time can be assessed, and the influence of transmitted voltage and MB dosage in each bolus injection can be explored. Our findings underscore the importance of optimal timing for data acquisition and highlight the necessity of real-time MB monitoring for successful ULM implementation. This research aims to provide insights into optimizing data acquisition for clinical implementation of ULM to increase the success rate of the ULM scanning, and potentially facilitating the future development of guidelines for standardizing ULM in clinical settings, paving the way for large-scale clinical use of the super-resolution ultrasound microvessel imaging techniques.

## Materials and Methods

To quantitatively investigate MB signal changes over time, ultrasound data were captured in a packet-by-packet manner following each bolus injection, covering the entire perfusion time course from a pig and a human in vivo. Quantitative indices of MB signal quality for localization were derived from each data packet, and their temporal changes were analyzed to facilitate the identification of optimal data acquisition timing for the clinical implementation of ULM. The effects of transmitted voltage and MB dosage on temporal MB signal changes were also evaluated.

*Experimental protocol*

1. Ultrasound data acquisition protocol

A Verasonics Vantage ultrasound system (Verasonics Inc., Kirkland, WA) equipped with a GE 9L linear array probe (GE Healthcare, Wauwatosa, WI, USA) was used in this study. A 6-angle compounding plane wave imaging (angle increment of 2°, center frequency of 5.2 MHz, and pulse length of 1 cycle) sequence was implemented with a post-compounding frame rate of 500 Hz for data acquisition. A packet of post-compound ultrasound in-phase quadrature (IQ) data with a packet size of 70 frames was captured, processed, and saved to the hard drive within every 1-3 seconds after MB injection. The time interval between adjacent packets may vary depending on the size of the beamformed field-of-view (FOV). The spatial pixel size of the saved IQ data was 0.5 wavelength in axial direction and 1.0 wavelength in lateral direction. Each packet of data was processed via an adaptive EVD-based clutter filter accelerated by graphics processing unit (GPU) [60] and the cine-loops of the flowing MB images were quickly displayed to enable a semi-real time monitoring of MBs during data acquisition, which may provide helpful feedback for the sonographer to adjust image plane accordingly during long periods of hand-held scanning in in-human study. With such sequence setup, a series of data packets will be captured to study the changes in MB signals over time after MB injection. We tested the influence of injected MB dosage and transmitted voltage on the dynamic changes of MB signal over time.

2. Ultrasound data processing

Each packet of acquired ultrasound IQ data was first filtered by an adaptive spatial-temporal clutter filter to extract the MB signal [60]. Data packet with large motions can be identified and excluded from analysis based on the eigenvalue distribution. The misalignment between data frames caused by the strong tissue motion can result in significant rank inflation, leading to a more flatten eigenvalue curve. Specifically, in this study, data packet with the $10^{th}$ largest eigenvalue above -30 dB compared to the largest eigenvalue is considered with large tissue motion and was excluded. Analysis of the remaining data packet by packet will provide the MB signal changes over the time course after MB injection. All data processing was implemented in MATLAB R2021a (MathWorks Inc., Natick, MA).

3. Quantitative assessment of MB signal

Robust MB localization using ULM is generally associated with high MB SNR, less MB overlapping or distortion, and sufficient number of localizable MBs in the limited time frame, which in combination provides a good imaging condition and timing for data acquisition in clinical scanning scenarios. Therefore, we mainly considered the following three aspects of the clutter-filtered MB data for the assessment of MB signal quality for robust MB localization and ULM in this study:

1) MB signal intensity and its relationship to noise level. The overall MB signal intensity is calculated as the mean power of the MB signal within the ROI averaged along slow-time direction for each data packet, and analysis of all the data packets will provide a MB signal intensity-time curve. This overall MB signal intensity is dominated by the MB concentration, i.e., the number of MBs within the ROI, without necessarily reflecting the SNR of individual MBs. In an attempt to analyze the SNR of individual MBs, we first localized the MBs following the ULM signal processing procedure [6]. More specifically, the clutter filtered MB data were first equalized by dividing them by a measured noise intensity profile to equalize the noise distribution throughout the image FOV [61]. An intensity threshold was then applied to the MB envelope data to remove the noisy background, followed by 2D normalized cross-correlation between the MB data and the predefined point-spread function (PSF) specific to the imaging system and settings derived from a multivariate Gaussian function. The normalized cross-correlation map was further thresholded to remove pixels with a correlation coefficient <0.3, and the regional peaks of which were identified as the MB centroids. With MB localized, the signal intensity at the MB centroid positions is assumed to be $S(i) = S_{MB}(i) + n(i)$, where $S_{MB}(i)$ is the pure signal intensity of $i^{th}$ MB, the $n(i)$ is the additive random noise of this pixel position, and $i$ is the number order of the MBs being localized. The average signal power of the $S(i)$ can thus be estimated as $\sum_i S^2(i)/N \approx \sum_i S_{MB}^2(i)/N + \sum_i n^2(i)/N$, assuming that MB signal $S_{MB}(i)$ and random noise $n(i)$ are uncorrelated and thus cross-term is omitted with zero expectation. Here $N$ is the total number of MBs localized in each packet of data. Therefore, the average SNR of the MB centroids for those localizable MBs can be approximated as $SNR = 10 \log_{10} \frac{\sum_i S_{MB}^2(i)/N}{\sum_i n^2(i)/N} \approx 10 \log_{10} \frac{\sum_i S^2(i) - \sum_i n^2(i)}{\sum_i n^2(i)}$, where $n(i)$ is the equalized noise intensity with an expectation of $E(\sum_i n^2(i)/N) = 1$.

2) Correlation of the MB signal with the PSF. Robust MB localization typically requires a well-defined profile of the MB signals that closely resembles the PSF of the imaging system. In addition to the SNR, the quality of MB profile is directly related to the separability or sparsity of the MBs, as excessive overlapping or distortion can significantly comprise the localizability of the MBs [47]. Factors such as imaging artifacts from phase aberration, reverberation, and attenuation can further degrade MB signal quality. In the localization processing procedure mentioned above, the MBs are localized as the regional maxima of the normalized cross-correlation coefficient (NCC) map, which is indeed an indication of the similarity between the MB signals and the given PSF. A higher value of NCC maxima is associated with a better MB signal quality for better localization, while a low value is related to degraded signal quality which can result from factors such as high MB concentration, low SNR, image artifacts, or a combination of these factors. Therefore, we used the NCC values at the positions of those localizable MBs within the ROI as the primary index for assessing the overall MB signal quality for localization in this study.

3) The number of MBs effectively identified and localized. For optimal ULM performance, it is crucial to have not only a sparse MB distribution and high MB SNR for precise localization but also a sufficient number of localized MBs within a short period to ensure dense microvasculature reconstruction.

Accordingly, we measured the number of localized MBs per frame per unit area and evaluated its changes over time following a bolus MB injection. An effective data acquisition timing post-MB injection aims to achieve high quality MB localization while maximizing the number of localized MBs to ensure efficient data acquisition within the limited time frame available in clinical applications.

4. Continuous data acquisition for clinical ULM

The packet-by-packet data acquisition scheme described above was specifically designed to study the changes in MB signal characteristics over the time course of bolus MB injection, to enhance our understanding of MB dynamics and provides insights into optimal imaging timing. However, for practical ULM implementation in clinical settings, it is preferable to continuously acquire sufficient data within a short time frame, ideally during a single breath-hold, to minimize tissue motion and misalignment artifacts [45, 46]. To address this need, we designed a continuous data acquisition sequence using the same Verasonics scanner and probe for actual ULM implementation: 1) A real-time MB imaging enabled by GPU-accelerated adaptive EVD-based clutter filtering is used to provide a real-time display of MB signals, allowing the sonographer to ensure a proper timing and imaging settings; 2) Once an optimal time window is identified, a long packet of ultrafast ultrasound IQ data with a frame rate of 500 Hz (4800 frames, corresponding to 9.6 s) is captured in one acquisition. An 8-angle compounding plane wave imaging ($2^o$ angle increment, 5.2 MHz center frequency, 1 pulse cycle) was used with the same beamformed pixel size as above. In the patient, the participant was asked to pause the breathing during data acquisition. The acquired data were processed offline to reconstruct ULM images with a pixel size of 50 μm in both spatial dimensions in MATLAB [6]. To improve MB data separability, the original MB data was first split into two subsets corresponding to the positive and negative Doppler frequencies, respectively, before localization and MB tracking [5].

*Pig study*

The animal study was approved by the Institutional Animal Care and Use Committee (IACUC). A domestic pig (aged 10 weeks, 56 kg) was anesthetized using intramuscular Telazol (5 mg/kg) and Xylazine (2 mg/kg), with anesthesia maintained through intravenous administration of ketamine (11 mg/kg/hr) and Xylazine (1.8 mg/kg/hr). The pig was intubated and mechanically ventilated with room air, followed by placement of intravascular catheters in the carotid artery and external jugular vein. Given the potential for acute allergic reactions to ultrasound contrast agents in pigs, intravenous injections of diphenhydramine (1 mg/kg) and dexamethasone (2 mg) were administered 30–60 minutes before MB injection to prevent adverse reactions [62-64]. Blood pressure was monitored through the arterial catheter, oxygen levels were assessed using an optical pulse oximeter attached to the pig's ear, and heart rate was tracked using ECG leads throughout the study. The pig was placed in a lateral position, and fur at the ultrasound scanning site was removed with a shaver. The kidney was scanned along a longitudinal view, and the ultrasound probe was clamped during data acquisition to minimize motion. Both packet-by-packet and continuous data acquisition sequences, as described above, were used to capture data following manual bolus injections of Definity MB suspension (Lantheus Inc., MA) through the jugular vein catheter, each followed by a flush of saline solution. MB signal changes under different transmitted voltage settings (5 V, 10 V, and 15 V) and MB injection dosages (1 mL and 2 mL) were evaluated.

*Human study*

The human study was approved by the Institutional Review Board (IRB) of Mayo Clinic, and written informed consent was obtained from the participant. Data was acquired from the kidney (longitudinal view) of a healthy volunteer (25-year-old female) in a lateral decubitus position using handheld scanning during each manual intravenous bolus injection session. For packet-by-packet data acquisition, the participant was instructed to breathe normally and avoid significant movement. This acquisition section lasted several minutes until the residual MBs were largely cleared, during which the sonographer adjusted the imaging plane as needed to maintain a scanning plane as consistent as possible. To expedite MB clearance between bolus injections and ensure minimal residual MBs for re-injection, scanning at higher MI settings, such as B-mode or color Doppler mode, might be performed as needed [58]. For continuous data acquisition section, real-time MB imaging mode was used after each MB injection to monitor MB concentration, and the participant was asked to hold the breath during the 9.6 seconds data acquisition. Different transmitted voltages (5 V, 10 V) and MB injection dosages (0.5 mL, 1 mL, and 2 mL) were tested to assess their impact on MB signal changes post-injection.

RESULTS

*Pig study*

1. MB Signal Change Over Time after Bolus Injection

Quantitative indices were derived from the ROI in the kidney cortex delineated in Fig. 1c (second column) for each data packet collected after a bolus MB injection. The temporal changes in these indices for a 1 mL bolus injection at a transmitted voltage of 5 V are shown in Fig. 1b. Representative MB images and ULM images from selected data packets (70 frames, 500 Hz, equating to 0.14 s of data) at different post-injection time points are displayed in Fig. 1c. The signal intensity curve, similar to the time-intensity curve in conventional CEUS, exhibits a rapid increase during the wash-in phase, followed by a decrease in the wash-out phase (Fig. 1b, upper left). Unlike the gradual decline observed in humans, the wash-out phase in pigs is characterized by a rapid intensity drop, reaching a bottom plateau within ~10 seconds (Fig. 1b). The SNR of those localizable MBs follows a similar pattern, with a fast increase to the peak (~35 dB) and then a rapid decrease to a lower and relative constant SNR level when MB is sparse (Fig. 1b, upper right).

The NCC is a composite quality index reflecting both SNR and MB concentration. NCC is expected to be high as MB first appear, and then rapidly drops to a minimum at peak MB concentration (when MBs are densely overlapping), followed by an increase to a plateau as MB is sparse again (Fig. 1b, bottom left; Fig. 1c). At the NCC minimum, overlapping MB signals hinder MB isolation, leading to noisy, short, and inconsistent tracks in ULM images (Fig. 1c, 6 s). As MB concentration declines passed peak, the MB signals become sparse and localizable (Fig. 1c, 13–60 s), yielding an increased NCC value and more consistent and continuous tracks in ULM images (Fig. 1b). The change of MB counts over post-injection time is shown in Fig. 1b (bottom right). Similarly, the localized MB number undergoes a fast increase to a peak, followed by a decline during wash-out phase. Note that the time-MB number curve is saturated as at peak MB concentration the localization is inaccurate and the number of detected MBs is expected to be highly underestimated. As NCC increases after peak, the number of detectable MBs drops fast to a turning point in about 10s, and afterwards the number of MBs become too sparse to form meaningful vessel structure within the short period of time for each data packet (Fig. 1c, last column).

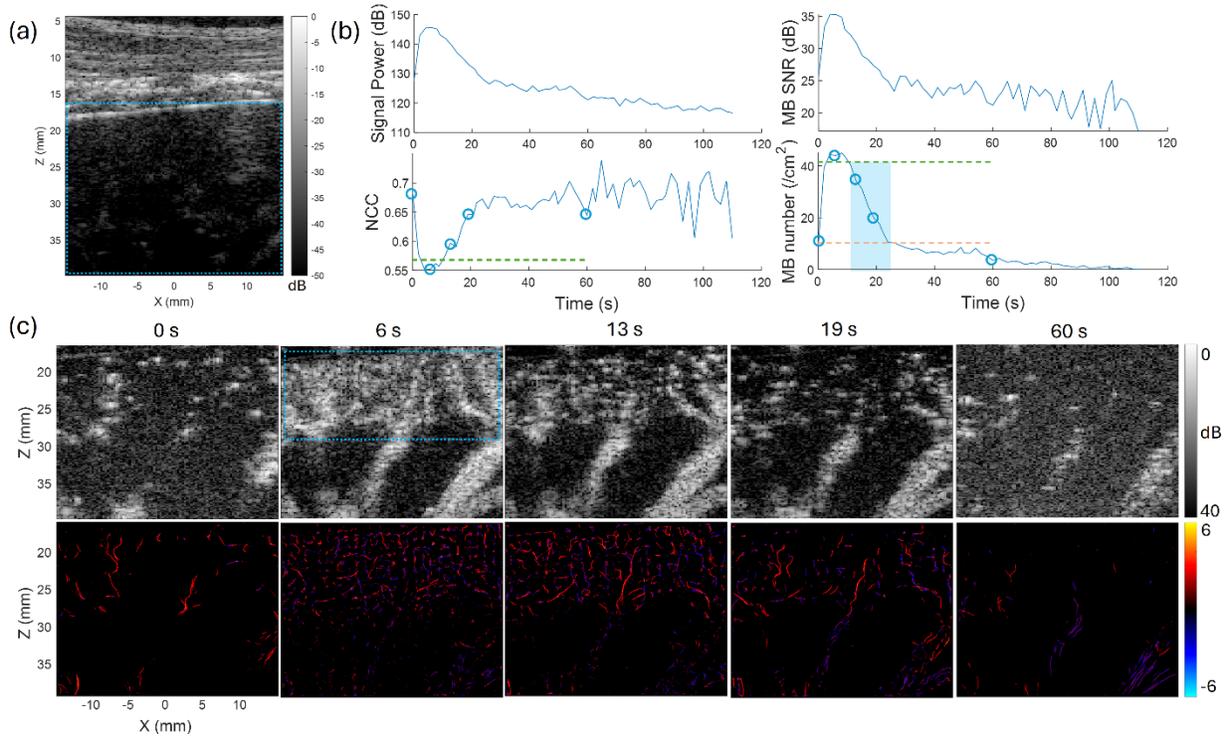

Fig. 1. (a) B-mode ultrasound image of the pig kidney. (b) Temporal changes in quantitative indices of MB signals, including signal power, individual MB SNR, NCC, and the number of MBs within the FOV, following a 1 mL bolus injection at a transmitted voltage of 5 V. The green dashed line in the time-MB number plot marks the approximate saturation threshold, above which the MB concentration is too high for accurate MB localization (corresponding to the NCC values below 0.57 in the time-NCC plot in this case). The horizontal orange dashed line indicates an approximate threshold below which the MB counts may be considered insufficient for efficient MB filling in a short period of time. The shadowed region between these thresholds represents an optimal imaging window where localization quality (high NCC) and MB counts are well-balanced. (c) Representative MB images and ULM density maps at different post-injection time points, indicated by the circles on the time-NCC and time-MB number curves. Each ULM density map was derived from the corresponding data packet at the indicated time point (70 frames, 500 Hz, equating to 0.14 s of data), with red and blue colors representing MB flow directions (upward and downward, respectively). The dynamic range is compressed by taking the square root.

A good timing for data acquisition of the pig kidney may fall in a range when the MB concentration is below localization saturation (allows for localization of individual MBs, approximately indicated by the green dashed line in Fig. 1b) and the number of localized MBs above an acceptable threshold (allows for fast vessel filling, approximately indicated by the orange dashed line in Fig. 1b). For this case, the ULM images accumulated from the available 4 data packets (4×70 frames, 0.56 s of data) captured around concentration peaks with NCC < 0.57 (indicated by the green line in Fig. 1b) is shown in Figs. 2a-2b, indicating a noisier MB tracks and less clean vessel structures. In contrast, the ULM images accumulated from the available 8 data packets (8×70 frames, 1.12 s of data) during the wash-out phase with NCC > 0.57 (shadowed region indicated in Fig. 1b) reveal a clearer MB tracks and microvasculature (Fig. 2c, 2d), indicating the importance of good timing for successful ULM imaging, particularly in the pig kidney model with rapid MB clearance.

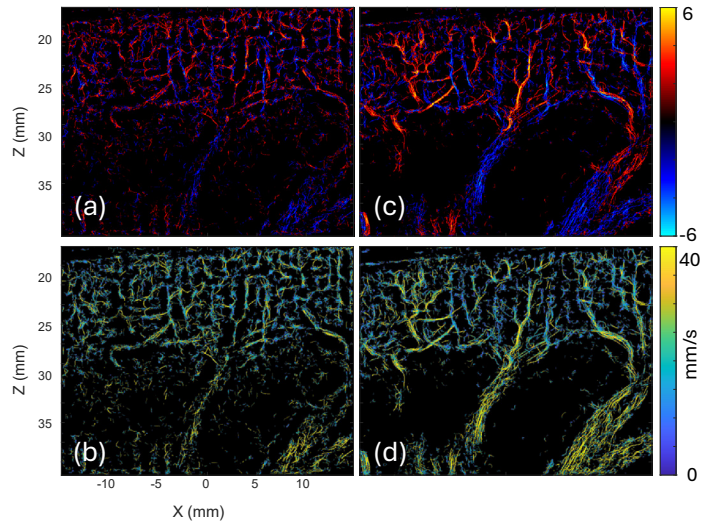

Fig. 2. (a) ULM density image and (b) velocity image from the pig study reconstructed from 4 data packets (4 × 70 frames, 0.56 s of data) acquired at peak MB concentration with NCC < 0.57 (below the green dashed line in the time-NCC plot in Fig. 1b). (c) ULM density image and (d) velocity image reconstructed from 8 data packets (8 × 70 frames, 1.12 s of data) acquired within the shadowed time window indicated in Fig. 1b, where NCC > 0.57.

2. Influence of Bolus Injection Dosage

The temporal changes in MB signal characteristics following 1 mL and 2 mL bolus injections of MB suspension, scanned at a fixed transmitted voltage of 15 V, are shown in Fig. 3. Both dosages exhibit similar temporal patterns across all indices; however, the 2 mL injection results in higher signal power and MB SNR at the peak phase, as well as a broader peak phase, as expected (Fig. 3a, 3b). During the peak concentration, the 2 mL injection also shows a prolonged phase of lower NCC values, consistent with a higher degree of MB overlap at higher dosages (Fig. 3c). The higher dosage delays the onset of the wash-out phase and extends the saturation period, as seen in the time-MB number curve (Fig. 3d). As a result, the optimal timing in the wash-out phase (when NCC starts to increase, and MB count starts to reduce) is shifted to later time points for the 2 mL dose. Despite this delay, a similar duration of about 10 s of the optimal imaging window (time range from saturation threshold to MB number threshold) is observed for both dosages. In the later phase (>40 seconds post-injection), when MB concentration becomes sufficiently sparse, all quantitative indices converge to similar levels for both dosages. Therefore, increasing the dosage from 1 ml to 2 ml does not significantly extend the duration of the optimal imaging window for the pig kidney in this case, but delays the onset of the optimal timing, implying the potential need for dose-specific adjustments to the imaging timing for robust ULM in clinical settings.

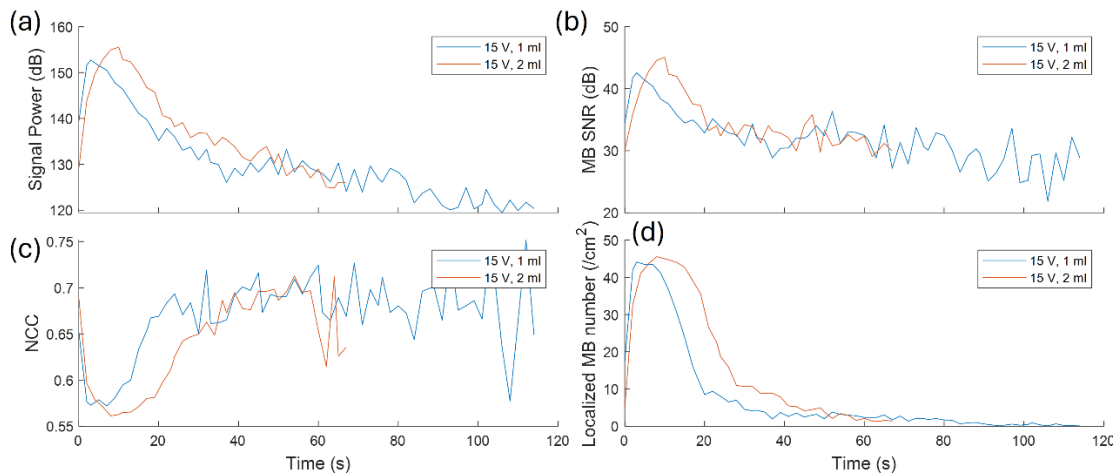

Fig. 3. Temporal changes in MB signal characteristics from the pig study following bolus injections of 1 mL and 2 mL. MB signal characteristics include (a) signal power, (b) individual MB SNR, (c) NCC, and (d) the number of MBs within the FOV, and acoustic transmitted voltage was 15 V.

3. Influence of Acoustic Power

The impact of transmitted voltage on MB signal characteristics for a fixed 1 ml bolus injection is illustrated in Fig. 4. All quantitative indices follow similar temporal trends across the three voltage settings (5 V, 10 V and 15 V). Higher transmitted voltages are associated with an increased MB signal intensity and MB SNR, with 15 V condition yielding a consistent improvement throughout the time course compared to the 10 V and 5 V conditions (Fig. 4a, 4b). NCC values show a minor improvement at higher voltages (10 V and 15 V) during the peak concentration phase, indicating a slight localization improvement under high acoustic power when MB overlap dominates (Fig. 4c). However, the differences are less pronounced, and NCC values converge to similar levels during the later phase when MB concentrations become sparse (> 20 s post-injection, Fig. 4c). The number of detected MBs shows limited changes among voltage settings, though the MB count drops slightly slower for 5 V (indicated by the arrow in Fig. 4d) compared to the 10 V and 15 V transmissions, potentially due to occurrence of MB destruction at elevated voltages.

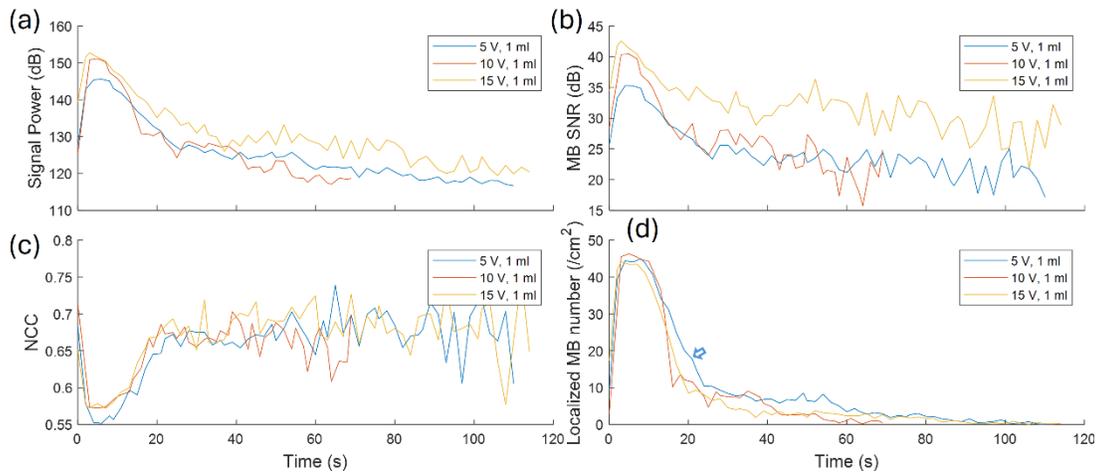

Fig. 4. Temporal changes in MB signal characteristics from the pig study following 1 mL bolus injections under three different transmitted voltage settings (5 V, 10 V, and 15 V). The MB signal characteristics include (a) signal power, (b) individual MB SNR, (c) NCC, and (d) the number of MBs within the FOV.

4. Continuous Data Acquisition for ULM

To accommodate clinical scenarios, we programmed to continuously acquire 9.6 s of data, designed to fit within a single breath-hold for in vivo applications. Based on the MB signal characteristics observed during bolus injections, the optimal timing for ULM data acquisition in the pig kidney occurs during the wash-out phase, when MB concentrations drop below saturation for localization but before MBs are largely cleared out. Three datasets were acquired at either the dropping phase or near the peak phase in separate bolus injections (1 mL dosage, different transmitted voltages). Similar to the above analysis, the changes of NCC and MB number within this 9.6s time frame was also calculated in a sliding window manner, as shown in Fig. 5. For fair comparison, the same amount of 3.6 s of data with a similar localization quality (NCC approximately in a range of 0.6-0.65) and MB counts (approximately in a range of 40-25 MBs/cm$^2$) from the dropping phase indicated by the shadowed regions was used to generate the ULM images (Fig. 5). All three datasets yielded ULM images with well-organized and consistent microvascular structures in the cortical regions of the kidney, indicating good repeatability of scanning enabled by just a few of seconds of data acquisition within optimal timing window. However, for the dataset acquired at 15 V transmitted voltage, the visualization of finer vessels, particularly beneath the kidney capsule, appeared slightly reduced, potentially due to the MB destruction at higher transmitted voltages.

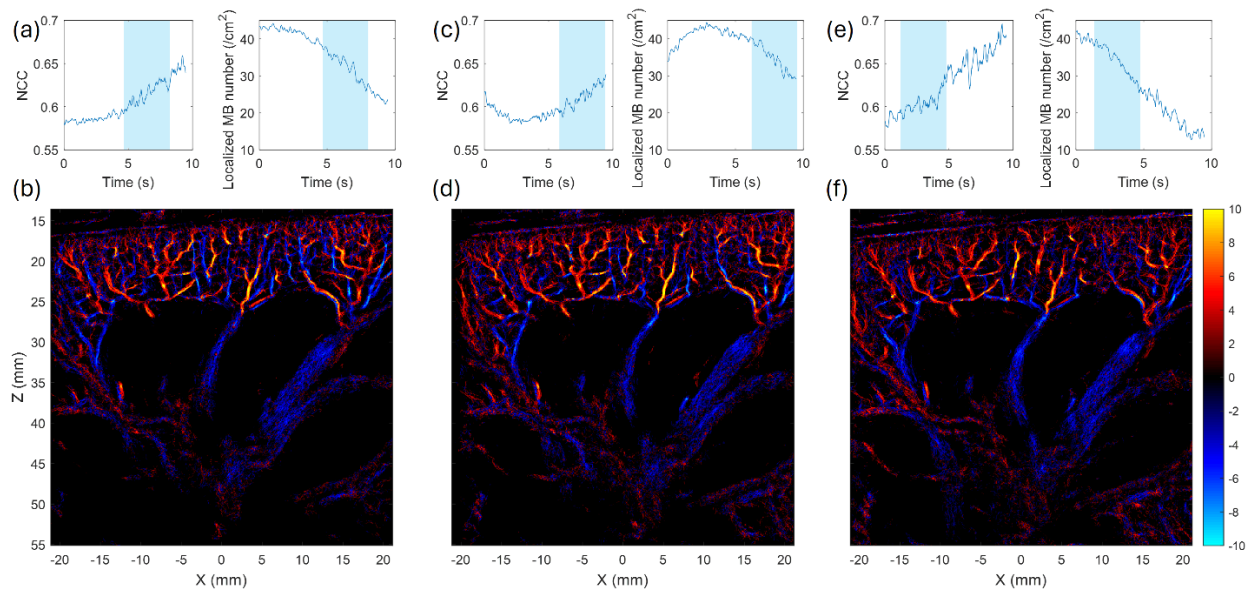

Fig. 5. The temporal changes of NCC and MB counts within the 9.6 s time frame of data acquisition in separate bolus injections (1 mL each) at transmitted voltages of (a) 5 V, (c) 10 V and (e) 15 V from the pig study. Panels (b), (d) and (f) are the corresponding ULM images, reconstructed using the same amount of 3.6 s of data from each acquisition, selected to ensure similar localization quality (NCC approximately 0.6–0.65) and MB counts (approximately 40–25 MBs/cm²), as indicated by the shadowed regions.

*In human study*

1. MB Signal Change Over Time After Bolus Injection

Figure 6 illustrates the temporal changes in MB signal characteristics following a 1 mL bolus injection at a transmitted voltage of 10 V. Unlike pig study, the wash-out phase in humans exhibits a significantly slower decline in MB concentration. For instance, it takes over 2 minutes for the signal power to decline by 15 dB from its peak, whereas this drop occurs in approximately 10 seconds in pigs, reflecting a significant difference in physiological MB clearance rate. Similarly, NCC decreases to its minimum at peak MB concentration within about 10 seconds after MB injection, then gradually increases during the wash-out phase, indicating improving localization quality as MB signals become less overlapped. Representative MB images and corresponding ULM images reconstructed from the corresponding data packet (70 frames, equating to 0.14 s of data) at different time points (marked by circles on the time-NCC and time-MB number curves) are shown in Fig. 6c, revealing the visual evolution of MB concentration and MB tracks over time. The NCC increases gradually in the wash-out phase as the MBs get sparser. However, it is difficult to define a clear NCC threshold where MB quality is adequate for precise localization. Empirically, if an NCC of 0.57 is still chosen as the threshold in this case (indicated by the green dashed lines in Fig. 6b), and an MB detection threshold (the minimum acceptable MB counts, indicated by the green dashed lines in Fig. 6b) is set at 25% of the MB number detected at saturated level, a time window of approximately 110 seconds (from 44 to 154 seconds post-injection) for data acquisition can be identified, as indicated by the shadowed region in Fig. 6b. The slow decline of MB concentration during the wash-out phase thus provides a larger time frame (approximately 2 minutes in this case) for optimal data acquisition in clinical bolus injection settings. However, as NCC increases while MB counts decrease, there is a tradeoff between MB localization quality and detected MB density for ULM in clinical bolus injection scenarios.

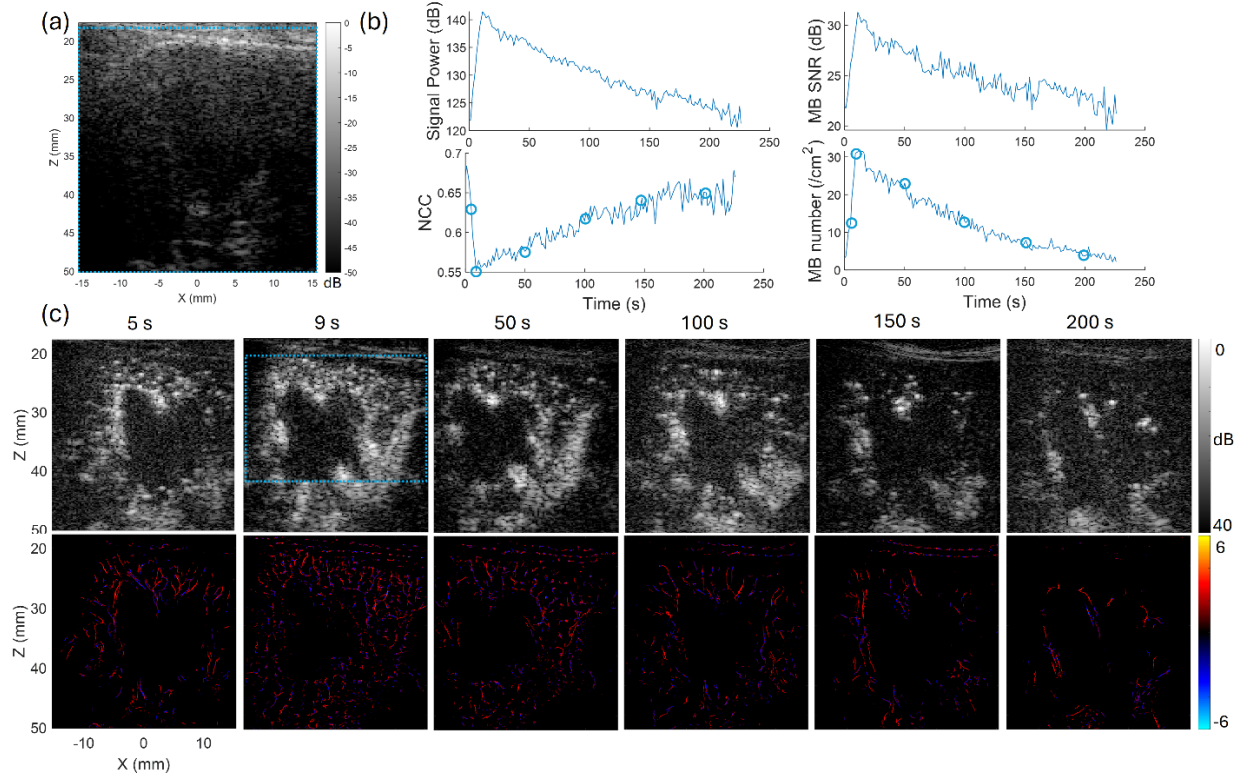

Fig. 6. (a) B-mode image of a human kidney. (b) Temporal changes in quantitative indices of MB signals, including signal power, individual MB SNR, NCC, and the number of MBs within the FOV (outlined by the rectangle in Fig. 6c), following a 1 mL bolus injection at a transmitted voltage of 10 V. The green dashed line in the time-MB number plot marks the empirical saturation threshold (corresponding to an NCC value of 0.57 in the time-NCC plot). The orange dashed line indicates an empirical MB detection threshold where the MB count drops to 25% of the saturated count, below which MB counts may be considered insufficient. The shadowed region between these thresholds represents an optimal ULM imaging window with a balance between localization quality and MB counts. (c) Representative MB images and ULM density maps at different post-injection time points, indicated by the circles on the time-NCC and time-MB number curves. Each ULM density map was derived from the data packet at the indicated time point (70 frames, 500 Hz, equating to 0.14 s of data), with red and blue colors representing upward and downward MB flow directions, respectively.

2. Influence of Bolus Injection Dosage

The temporal changes in MB signal characteristics for three bolus injection dosages (0.5 mL, 1 mL, and 1.5 mL) at a transmitted voltage of 10 V are shown in Fig. 7. Higher dosages result in elevated signal power and MB SNR, with a slightly delayed peak phase, consistent with observations from the pig study, and an extended wash-out phase compared to lower dosages. At the lower dosage (0.5 mL), NCC values tend to be higher at peak phase due to the reduced MB concentration and MB overlaps, potentially enabling localization even during the peak phase. However, as MB concentration steadily decreases during the wash-out phase, NCC values for the higher dosages (1 mL and 1.5 mL) gradually increase and eventually converge to the same level as those observed for the lower dosage in the later phase. The extended wash-out phase associated with higher injection dosages allows for an extended time window for data acquisition. However, the onset of the preferred timing for data acquisition is also delayed for higher dosages, revealing the importance of real-time monitoring of MB signals to ensure MB quality and density for robust ULM across different dosages.

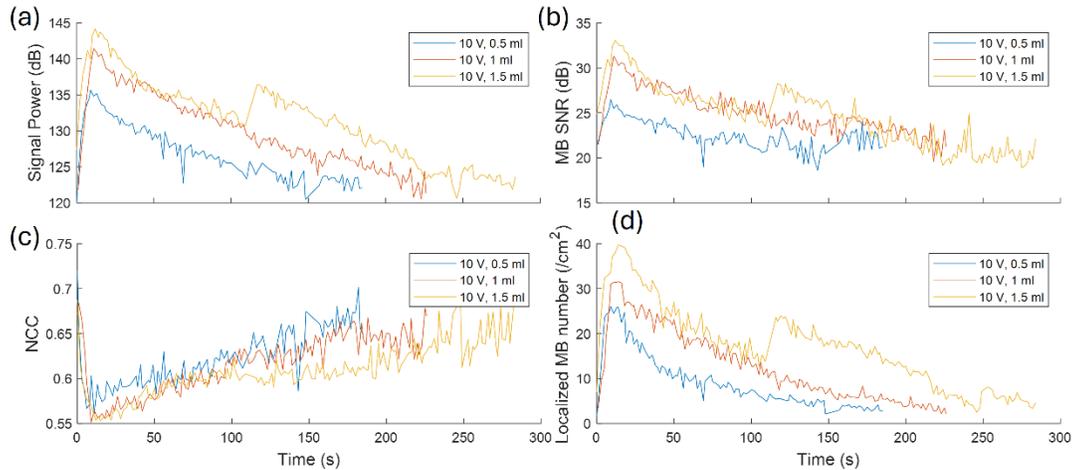

Fig. 7. Temporal changes in MB signal characteristics from the human study following bolus injections of 0.5 mL, 1 mL, and 1.5 mL at a transmitted voltage of 10 V. The MB signal characteristics include (a) signal power, (b) individual MB SNR, (c) NCC, and (d) the number of MBs within the FOV.

3. Influence of Acoustic Power

Figure 8 illustrates the effect of acoustic power on MB signal characteristics following 1 mL bolus injections under different transmitted voltage settings (5 V and 10 V). Higher transmitted voltages result in consistently increased signal intensity and SNR, as expected. However, no significant increase in NCC is shown, indicating that both acoustic power settings yield a similar quality in terms of MB localization. MB counts remain largely similar between the two voltage settings, which is expected as the MB dosage is constant. However, a slight decline in MB counts is observed at the higher transmitted voltage (10 V) during the later phase (>100 seconds post-injection, Fig. 8d), which may be attributed to MB destruction caused by elevated acoustic power.

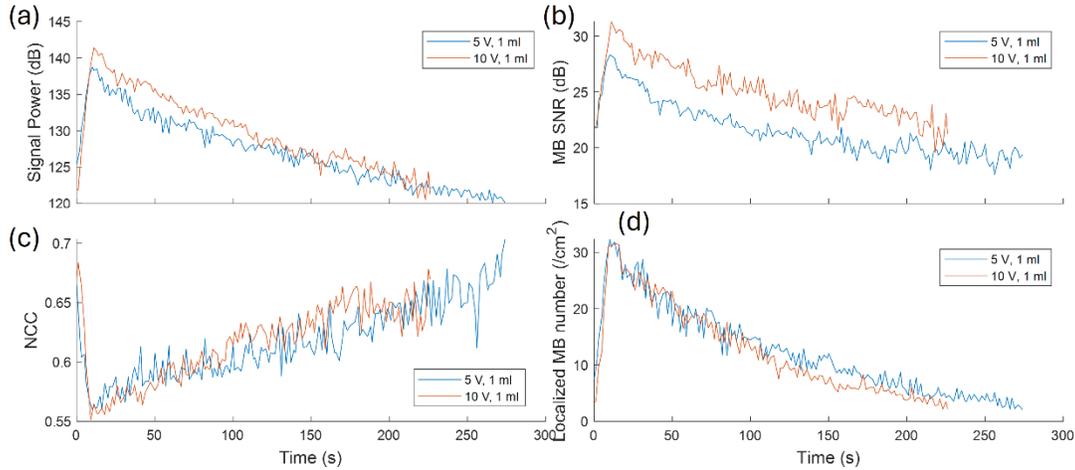

Fig.8. Temporal changes in MB signal characteristics from the human study following 1 mL bolus injections under different transmitted voltage settings (5 V and 10 V). The MB signal characteristics include (a) signal power, (b) individual MB SNR, (c) NCC, and (d) the number of MBs within the FOV.

4. Continuous Data Acquisition for ULM

For practical ULM implementation, 9.6 seconds of data were continuously captured in a single acquisition within a breath hold. Two datasets were acquired from separate 1 mL injections at a transmitted voltage of 10 V, both during the wash-out phase within the identified optimal data acquisition window: one at approximately 50 seconds post-injection and the other at around 150 seconds. The temporal changes in NCC and MB counts during these 9.6-second acquisitions are shown in Fig. 9a and 9e. As expected, the changes in NCC and MB counts during this short acquisition period were relatively minor compared to the much longer overall time scale of the wash-out phase (approximately 4 minutes, as shown in Figs. 6–8). Slight oscillations observed in the NCC and MB count curves (Fig. 9a and 9e) correspond to the patient's heart rate, reflecting cyclic perturbations in MB density and localization quality. Note that MB counts per unit area are influenced by the imaging plane and ROI selection, so the absolute MB counts for the two datasets with different imaging planes may not be directly compared.

Both datasets produced high-quality ULM images, revealing well-defined vascular networks within the kidney cortex using data from a single breath hold (Fig. 9). However, as expected, the second dataset, acquired later in the wash-out phase, exhibited higher NCC values (~0.65) due to lower MB density (Fig. 9f), while the first dataset reflected a denser MB distribution, as indicated in the B-mode image of MB signals (Fig. 9b, NCC ~0.61). At higher MB concentrations (Fig. 9c–9d), vessel structures appeared to fill more rapidly and seemed slightly denser. However, noisier MB tracks were observed, likely due to the relatively lower MB localization quality (lower NCC values). In contrast, at lower MB concentrations (Fig. 9g–9h), MB tracks and reconstructed microvascular structures appeared sharper, less noisy, with improved velocity estimations, particularly in larger vessels associated with the better MB localization (higher NCC values), again revealing a tradeoff between MB localization quality and vessel filling speed.

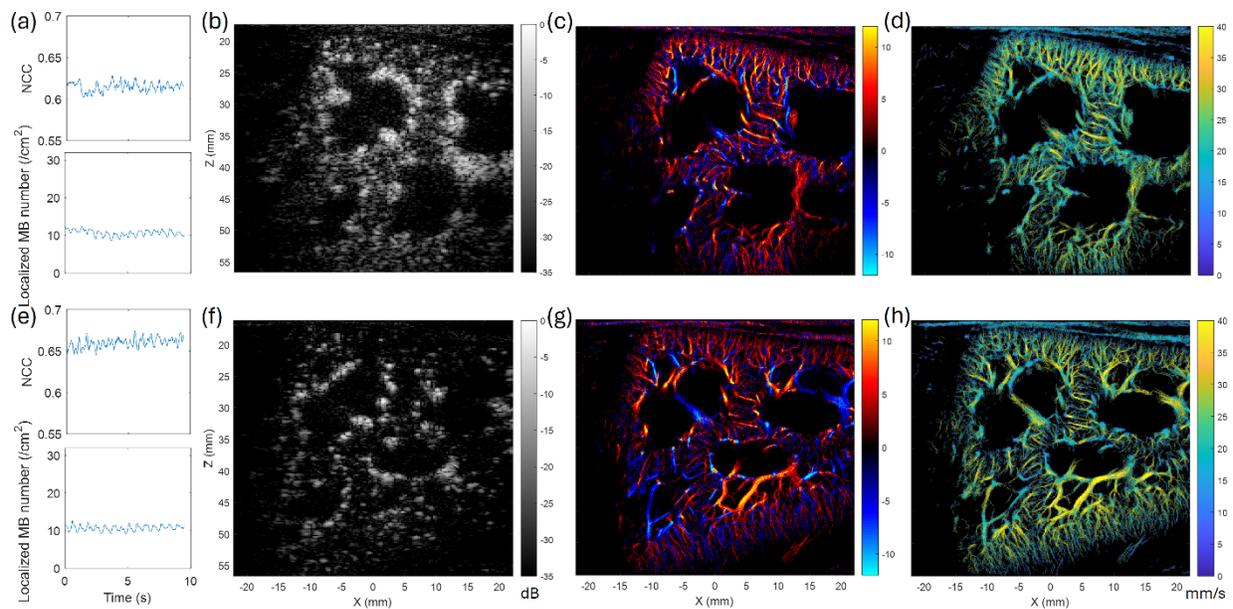

Fig. 9. Temporal changes in NCC and MB counts within the 9.6-second data acquisition window at approximately 50 seconds after a 1 mL bolus injection at a transmitted voltage of 10 V from the human study. (b) Example B-mode image of the MB data from this acquisition. (c) Corresponding reconstructed ULM density image. (d) Corresponding ULM velocity image. (e) Temporal changes in NCC and MB counts within the 9.6-second data acquisition window at approximately 150 seconds after another 1 mL bolus injection at a transmitted voltage of 10 V. (f) Example B-mode image of the MB data from this acquisition. (g) Corresponding reconstructed ULM density image. (h) Corresponding ULM velocity image.

## DISCUSSIONS

In this study, we investigated the temporal changes in MB signals during bolus injections in both pigs and humans in vivo in an attempt to facilitate optimal data acquisition for robust ULM implementation in clinical settings. Since ULM relies on the accurate localization of spatially isolated MBs, the dosage of the bolus injection and the timing of data acquisition post-injection are critical. Quantitative indices, including MB signal power, SNR, and NCC, were used to assess MB signal quality for localization throughout the perfusion course. Beyond MB data quality, an optimal dataset for localization requires a sufficient number of localized MBs to ensure rapid vessel filling within a short period of time suitable for in vivo applications. The inherent tradeoff between MB concentration and localization quality was studied to guide acquisition timing under varying transmitted voltages and MB injection dosages. Using this quantitative analysis of MB changes in bolus injection, we were able to identify appropriate time windows for data acquisition in both pig and human studies. In the pig study, data acquisition during the optimal timing window produced repeatable high-quality ULM images within just a few seconds, despite the challenges of rapid MB clearance. Similarly, robust ULM was successfully achieved in humans using a short period of data captured in one breath hold, and a tradeoff between MB density and localization quality at different wash-out phases was revealed. Therefore, the study on the temporal change of MB signal in bolus injection provides important insights for achieving robust and reproducible ULM in clinical settings and has the potential to facilitate standardization and widespread clinical translation of ULM.

In pigs, the rapid wash-out phase associated with high MB clearance rates resulted in a narrow optimal acquisition time window of about 10 seconds (the rapid dropping phase in the time-MB count curve). Real-time MB signal monitoring post-injection is therefore necessary to precisely identify the optimal timing and initiate data acquisition for successful ULM implementation. In contrast, a much slower wash-out phase was observed in humans, providing an extended time frame for data acquisition, with an optimal imaging window of approximately 1–2 minutes for a 1 mL bolus injection (Figs. 6–8). This prolonged window allows greater flexibility in clinical ULM scans and even enables multiple acquisitions within a single bolus injection, but also introduces an inherent tradeoff between localization quality (NCC) and MB density. Higher MB concentrations during the early wash-out phase facilitate faster vessel filling but compromise localization quality, while lower MB concentrations during the later wash-out phase improve localization accuracy but require longer acquisition times for complete vessel filling. Nevertheless, even in later phase at around 150 seconds post-injection, ULM data acquired within a single breath hold produced well-defined microvascular networks, demonstrating the feasibility of ULM in clinical practice. The timing and optimal scanning window may vary depending on the target organ, imaging FOV, or disease status of the organ [65]. Real-time MB monitoring or signal quality evaluation would be beneficial for identifying the optimal scanning time, adjusting imaging planes, and modifying imaging parameters on a case-by-case basis to achieve robust in vivo ULM.

Although CEUS may provide recommended dosages for scanning specific organs [58, 59], these dosages may not be necessarily applicable to ULM, which uniquely relies on MB sparsity. Consequently, ULM is expected to require lower MB dosages than those used in CEUS. Related to the dosage, the timing of ULM acquisition is critical, as MB concentration varies over time following a typical time-intensity curve. Nevertheless, even at higher dosages, a suitable time window for ULM can still be identified during the wash-out phase. It should be noticed that higher dosages can delay the peak concentration and extend the wash-out phase (Figs. 3, 7), necessitating adjustments to the timing of data acquisition for optimal ULM performance. Real-time monitoring of MB quality and concentration is therefore essential for ensuring accurate timing and imaging conditions. Appropriate acoustic output is also beneficial for maintaining MB signal quality while minimizing MB destruction. Higher acoustic outputs were associated with enhanced MB signal intensity and SNR in both studies (Figs. 4, 8) and can improve penetration for deeper MBs. However, the change of MB localization quality in terms of NCC was shown to be relatively modest by changing the voltage from 5 v to 10 v or 15 v in this study. This may be due to that the NCC values depend not only on SNR but also on MB signal fidelity, concentration, and morphology, leading to a similar localization quality for the range of acoustic powers investigated in this study. Additionally, higher voltages may be used cautiously, as they were associated with a more rapid decline in MB counts observed during the later wash-out phase, likely due to MB destruction.

While this study provides valuable insights into optimal data acquisition for in vivo ULM, several limitations warrant consideration. We focused on bolus injections, which are most clinically relevant. However, continuous infusion methods for MB administration are also an option, and the real-time monitoring and quantitative MB analysis strategies presented here could also be beneficial for optimizing infusion rates. While ULM primarily targets microvascular imaging, large vessels require different MB concentrations and scanning timings due to higher MB densities and faster flow rates, which make MB tracking more challenging [45]. Furthermore, MB changes in bolus injection can vary between organs and disease states, though this study focused solely on healthy kidney imaging. Nevertheless, the findings of this study emphasize the necessity of real-time MB monitoring to identify optimal scanning windows in practice for the application of ULM to different vascular systems. Although this study highlights the flexibility of the wash-out phase for identifying optimal imaging windows, the wash-out phase may also be leveraged for robust ULM, but the shorter time frame may introduce practical challenges for long data acquisition.

The field of super-resolution ultrasound microvascular imaging is rapidly evolving, with emerging methods aimed at improving MB localization and tracking or employing non-localization strategies for high MB concentrations [5, 12, 18, 48-50, 52, 54-57, 66]. The development of these methods holds promise to push boundary of optimal time window towards the peak MB concentration in clinical bolus injection and to further shorten the scanning time frame. Therefore, the optimal timing and imaging settings for data acquisition may still vary with actual algorithm used for super-resolution imaging. Additionally, higher frequency applications, aligned with CEUS practices, provide improved spatial resolution and may permit higher MB concentrations for MB localization. While curved-array transducers are advantageous for deeper penetration, this study used a linear-array transducer to achieve higher spatial resolution for MB detection, enabling the detection of more MBs within shorter acquisition time frames [47]. Ultimately, standardizing bolus injection protocols and imaging parameters across different clinical environments may be necessarily for ensuring robust and repeatable ULM performance and could facilitate broader adoption of ULM as a clinical tool for microvascular imaging.

## CONCLUSIONS

This study explored the temporal changes of MB signals for MB localization over the course of bolus injections in both pig model and in human to optimize data acquisition for robust ULM implementation in clinical settings. The effects of MB dosage and acoustic output on optimal scanning timing were also tested, and trade-offs between MB concentration and localization quality were revealed. Guided by these findings, robust ULM was achieved in both pig and human kidney using a short period of data, showing the feasibility of ULM in clinical practice. Therefore, the research offers insights into optimizing data acquisition for achieving consistent and reproducible ULM in clinical settings, paving the way for standardization and broader clinical applications of super-resolution ultrasound imaging.

REFERENCES

[1] O. Couture, B. Besson, G. Montaldo, M. Fink, and M. Tanter, "Microbubble ultrasound super-localization imaging (MUSLI)," in *2011 IEEE International Ultrasonics Symposium*, 2011, pp. 1285-1287.

[2] M. A. O'Reilly and K. Hynynen, "A super-resolution ultrasound method for brain vascular mapping," *Med. Phys.,* vol. 40, no. 11, p. 7, Nov 2013, Art. no. 110701.

[3] K. Christensen-Jeffries, R. J. Browning, M. X. Tang, C. Dunsby, and R. J. Eckersley, "In Vivo Acoustic Super-Resolution and Super-Resolved Velocity Mapping Using Microbubbles," *IEEE Trans. Med. Imaging,* vol. 34, no. 2, pp. 433-440, Feb 2015.

[4] C. Errico *et al.*, "Ultrafast ultrasound localization microscopy for deep super-resolution vascular imaging," *Nature,* vol. 527, no. 7579, pp. 499-+, Nov 2015.

[5] C. Huang *et al.*, "Short Acquisition Time Super-Resolution Ultrasound Microvessel Imaging via Microbubble Separation," *Sci Rep,* vol. 10, no. 1, p. 6007, 2020/04/07 2020.

[6] C. Huang *et al.*, "Super-resolution ultrasound localization microscopy based on a high frame-rate clinical ultrasound scanner: an in-human feasibility study," *Phys Med Biol,* vol. 66, no. 8, Apr 8 2021.

[7] F. L. Lin, S. E. Shelton, D. Espindola, J. D. Rojas, G. Pinton, and P. A. Dayton, "3-D Ultrasound Localization Microscopy for Identifying Microvascular Morphology Features of Tumor Angiogenesis at a Resolution Beyond the Diffraction Limit of Conventional Ultrasound," *Theranostics,* vol. 7, no. 1, pp. 196-204, 2017.

[8] P. F. Song *et al.*, "Improved Super-Resolution Ultrasound Microvessel Imaging With Spatiotemporal Nonlocal Means Filtering and Bipartite Graph-Based Microbubble Tracking," *IEEE Trans. Ultrason. Ferroelectr. Freq. Control,* vol. 65, no. 2, pp. 149-167, Feb 2018.

[9] O. M. Viessmann, R. J. Eckersley, K. Christensen-Jeffries, M. X. Tang, and C. Dunsby, "Acoustic super-resolution with ultrasound and microbubbles," *Phys. Med. Biol.,* vol. 58, no. 18, pp. 6447-6458, Sep 2013.

[10] J. Yu, L. Lavery, and K. Kim, "Super-resolution ultrasound imaging method for microvasculature in vivo with a high temporal accuracy," *Sci Rep,* vol. 8, p. 11, Sep 2018, Art. no. 13918.

[11] G. Zhang *et al.*, "Acoustic wave sparsely activated localization microscopy (AWSALM): Super-resolution ultrasound imaging using acoustic activation and deactivation of nanodroplets," *Appl. Phys. Lett.,* vol. 113, no. 1, p. 5, Jul 2018, Art. no. 014101.

[12] J. B. Zhang *et al.*, "Ultrasound Microvascular Imaging Based on Super-Resolution Radial Fluctuations," *J. Ultrasound Med.,* p. 10.

[13] J. Q. Zhu *et al.*, "3D Super-Resolution US Imaging of Rabbit Lymph Node Vasculature in Vivo by Using Microbubbles," *Radiology,* vol. 291, no. 3, pp. 642-650, Jun 2019.

[14] T. Opacic *et al.*, "Motion model ultrasound localization microscopy for preclinical and clinical multiparametric tumor characterization," *Nat. Commun.,* vol. 9, p. 13, Apr 2018, Art. no. 1527.

[15] A. Bar-Zion, O. Solomon, C. Tremblay-Darveau, D. Adam, and Y. C. Eldar, "SUSHI: Sparsity-Based Ultrasound Super-Resolution Hemodynamic Imaging," *IEEE Trans. Ultrason. Ferroelectr. Freq. Control,* vol. 65, no. 12, pp. 2365-2380, Dec 2018.

[16] S. B. Andersen *et al.*, "Super-Resolution Imaging with Ultrasound for Visualization of the Renal Microvasculature in Rats Before and After Renal Ischemia: A Pilot Study," *Diagnostics,* vol. 10, no. 11, p. 862, 2020.

[17] B. Heiles, A. Chavignon, V. Hingot, P. Lopez, E. Teston, and O. Couture, "Addendum: Performance benchmarking of microbubble-localization algorithms for ultrasound localization microscopy," *Nat Biomed Eng,* Oct 20 2023.

[18] R. J. G. van Sloun *et al.*, "Super-Resolution Ultrasound Localization Microscopy Through Deep Learning," *IEEE Trans Med Imaging,* vol. 40, no. 3, pp. 829-839, Mar 2021.

[19] C. Demené *et al.*, "Transcranial ultrafast ultrasound localization microscopy of brain vasculature in patients," *Nat Biomed Eng,* vol. 5, no. 3, pp. 219-228, Mar 2021.

[20] M. R. Lowerison *et al.*, "Aging-related cerebral microvascular changes visualized using ultrasound localization microscopy in the living mouse," *Sci Rep,* vol. 12, no. 1, p. 619, Jan 12 2022.

[21] O. Demeulenaere *et al.*, "Coronary Flow Assessment Using 3-Dimensional Ultrafast Ultrasound Localization Microscopy," *JACC Cardiovasc Imaging,* vol. 15, no. 7, pp. 1193-1208, Jul 2022.

[22] P. Cormier, J. Poree, C. Bourquin, and J. Provost, "Dynamic Myocardial Ultrasound Localization Angiography," *IEEE Trans Med Imaging,* vol. 40, no. 12, pp. 3379-3388, Dec 2021.

[23] S. Bodard *et al.*, "Ultrasound localization microscopy of the human kidney allograft on a clinical ultrasound scanner," *Kidney Int,* vol. 103, no. 5, pp. 930-935, May 2023.

[24] Z. Zhang, M. Hwang, T. J. Kilbaugh, A. Sridharan, and J. Katz, "Cerebral microcirculation mapped by echo particle tracking velocimetry quantifies the intracranial pressure and detects ischemia," *Nat Commun,* vol. 13, no. 1, p. 666, Feb 3 2022.

[25] J. R. McCall, F. Santibanez, H. Belgharbi, G. F. Pinton, and P. A. Dayton, "Non-invasive transcranial volumetric ultrasound localization microscopy of the rat brain with continuous, high volume-rate acquisition," *Theranostics,* vol. 13, no. 4, pp. 1235-1246, 2023.

[26] Q. Chen, J. Yu, B. M. Rush, S. D. Stocker, R. J. Tan, and K. Kim, "Ultrasound super-resolution imaging provides a noninvasive assessment of renal microvasculature changes during mouse acute kidney injury," *Kidney International,* vol. 98, no. 2, pp. 355-365, 2020/08/01/ 2020.

[27] X. Qian *et al.*, "Super-Resolution Ultrasound Localization Microscopy for Visualization of the Ocular Blood Flow," *IEEE Trans Biomed Eng,* vol. 69, no. 5, pp. 1585-1594, May 2022.

[28] J. Yan *et al.*, "Transthoracic ultrasound localization microscopy of myocardial vasculature in patients," *Nat Biomed Eng,* vol. 8, no. 6, pp. 689-700, Jun 2024.

[29] G. Tuccio, S. Afrakhteh, G. Iacca, and L. Demi, "Time Efficient Ultrasound Localization Microscopy Based on A Novel Radial Basis Function 2D Interpolation," *IEEE Trans Med Imaging,* vol. 43, no. 5, pp. 1690-1701, May 2024.

[30] J. N. Harmon, Z. Z. Khaing, J. E. Hyde, C. P. Hofstetter, C. Tremblay-Darveau, and M. F. Bruce, "Quantitative tissue perfusion imaging using nonlinear ultrasound localization microscopy," *Sci Rep,* vol. 12, no. 1, p. 21943, 2022/12/19 2022.

[31] T. M. Kierski *et al.*, "Superharmonic Ultrasound for Motion-Independent Localization Microscopy: Applications to Microvascular Imaging From Low to High Flow Rates," *IEEE Transactions on Ultrasonics, Ferroelectrics, and Frequency Control,* vol. 67, no. 5, pp. 957-967, 2020.

[32] K. Christensen-Jeffries *et al.*, "Super-Resolution Ultrasound Imaging," *Ultrasound Med. Biol.,* vol. 46, no. 4, pp. 865-891, Apr 2020.

[33] O. Couture, V. Hingot, B. Heiles, P. Muleki-Seya, and M. Tanter, "Ultrasound Localization Microscopy and Super-Resolution: A State of the Art," *IEEE Trans. Ultrason. Ferroelectr. Freq. Control,* vol. 65, no. 8, pp. 1304-1320, Aug 2018.

[34] S. Dencks *et al.*, "Clinical Pilot Application of Super-Resolution US Imaging in Breast Cancer," *IEEE Trans. Ultrason. Ferroelectr. Freq. Control,* vol. 66, no. 3, pp. 517-526, Mar 2019.

[35] J. Foiret, H. Zhang, T. Ilovitsh, L. Mahakian, S. Tam, and K. W. Ferrara, "Ultrasound localization microscopy to image and assess microvasculature in a rat kidney," *Sci Rep,* vol. 7, p. 12, Oct 2017, Art. no. 13662.

[36] D. Ghosh *et al.*, "Super-Resolution Ultrasound Imaging of Skeletal Muscle Microvascular Dysfunction in an Animal Model of Type 2 Diabetes," *J. Ultrasound Med.,* vol. 38, no. 10, pp. 2589-2599, Oct 2019.

[37] S. Harput *et al.*, "3-D Super-Resolution Ultrasound Imaging With a 2-D Sparse Array," *IEEE Trans. Ultrason. Ferroelectr. Freq. Control,* vol. 67, no. 2, pp. 269-277, Feb 2020.

[38] B. Heiles *et al.*, "Ultrafast 3D Ultrasound Localization Microscopy Using a 32 ×32 Matrix Array," *IEEE Trans. Med. Imaging,* vol. 38, no. 9, pp. 2005-2015, 2019.

[39] A. Chavignon, B. Heiles, V. Hingot, C. Orset, D. Vivien, and O. Couture, "3D Transcranial Ultrasound Localization Microscopy in the Rat Brain With a Multiplexed Matrix Probe," *IEEE Trans Biomed Eng,* vol. 69, no. 7, pp. 2132-2142, Jul 2022.

[40]	O. Demeulenaere et al., "In vivo whole brain microvascular imaging in mice using transcranial 3D Ultrasound Localization Microscopy," *EBioMedicine,* vol. 79, p. 103995, May 2022.

[41]	S. Harput et al., "3-D Super-Resolution Ultrasound Imaging With a 2-D Sparse Array," *IEEE Trans Ultrason Ferroelectr Freq Control,* vol. 67, no. 2, pp. 269-277, Feb 2020.

[42]	L. Qiu et al., "In vivo assessment of hypertensive nephrosclerosis using ultrasound localization microscopy," *Med Phys,* vol. 49, no. 4, pp. 2295-2308, Apr 2022.

[43]	H. Zhang et al., "Evaluation of Early Diabetic Kidney Disease Using Ultrasound Localization Microscopy: A Feasibility Study," *J Ultrasound Med,* vol. 42, no. 10, pp. 2277-2292, Oct 2023.

[44]	J. Yu, H. Dong, D. Ta, R. Xie, and K. Xu, "Super-resolution Ultrasound Microvascular Angiography for Spinal Cord Penumbra Imaging," *Ultrasound Med Biol,* vol. 49, no. 9, pp. 2140-2151, Sep 2023.

[45]	P. Song, J. M. Rubin, and M. R. Lowerison, "Super-resolution ultrasound microvascular imaging: Is it ready for clinical use?," *Z Med Phys,* vol. 33, no. 3, pp. 309-323, Aug 2023.

[46]	S. Dencks and G. Schmitz, "Ultrasound localization microscopy," *Zeitschrift für Medizinische Physik,* vol. 33, no. 3, pp. 292-308, 2023/08/01/ 2023.

[47]	M. Lerendegui, J. Yan, E. Stride, C. Dunsby, and M. X. Tang, "Understanding the effects of microbubble concentration on localization accuracy in super-resolution ultrasound imaging," *Phys Med Biol,* vol. 69, no. 11, May 20 2024.

[48]	X. Liu, T. Zhou, M. Lu, Y. Yang, Q. He, and J. Luo, "Deep Learning for Ultrasound Localization Microscopy," *IEEE Trans. Med. Imaging,* pp. 1-1, 2020.

[49]	Q. You et al., "Curvelet Transform-Based Sparsity Promoting Algorithm for Fast Ultrasound Localization Microscopy," *IEEE Trans Med Imaging,* vol. 41, no. 9, pp. 2385-2398, Sep 2022.

[50]	L. Milecki et al., "A Deep Learning Framework for Spatiotemporal Ultrasound Localization Microscopy," *IEEE Trans Med Imaging,* vol. 40, no. 5, pp. 1428-1437, May 2021.

[51]	A. Bar-Zion, C. Tremblay-Darveau, O. Solomon, D. Adam, and Y. C. Eldar, "Fast Vascular Ultrasound Imaging With Enhanced Spatial Resolution and Background Rejection," *IEEE Trans. Med. Imaging,* vol. 36, no. 1, pp. 169-180, Jan 2017.

[52] X. Chen, M. R. Lowerison, Z. Dong, N. V. Chandra Sekaran, D. A. Llano, and P. Song, "Localization Free Super-Resolution Microbubble Velocimetry Using a Long Short-Term Memory Neural Network," *IEEE Trans Med Imaging,* vol. 42, no. 8, pp. 2374-2385, Aug 2023.

[53] J. Arendt Jensen *et al.*, "Super-Resolution Ultrasound Imaging Using the Erythrocytes-Part I: Density Images," *IEEE Trans Ultrason Ferroelectr Freq Control,* vol. 71, no. 8, pp. 925-944, Aug 2024.

[54] J. H. Park, W. Choi, G. Y. Yoon, and S. J. Lee, "Deep Learning-Based Super-resolution Ultrasound Speckle Tracking Velocimetry," *Ultrasound Med Biol,* vol. 46, no. 3, pp. 598-609, Mar 2020.

[55] Y. Shin *et al.*, "Context-aware deep learning enables high-efficacy localization of high concentration microbubbles for super-resolution ultrasound localization microscopy," *Nat. Commun.,* vol. 15, no. 1, p. 2932, 2024/04/04 2024.

[56] G. Zhang *et al.*, "ULM-MbCNRT: In vivo Ultrafast Ultrasound Localization Microscopy by Combining Multi-branch CNN and Recursive Transformer," *IEEE Trans Ultrason Ferroelectr Freq Control,* vol. Pp, Apr 12 2024.

[57] G. Zhang *et al.*, "In vivo ultrasound localization microscopy for high-density microbubbles," *Ultrasonics,* vol. 143, p. 107410, 2024/09/01/ 2024.

[58] C. F. Dietrich *et al.*, "How to perform Contrast-Enhanced Ultrasound (CEUS)," *Ultrasound Int Open,* vol. 4, no. 1, pp. E2-e15, Jan 2018.

[59] C. Greis, "Technology overview: SonoVue (Bracco, Milan)," *Eur Radiol,* vol. 14 Suppl 8, pp. P11-5, Oct 2004.

[60] C. Huang, U.-W. Lok, J. Zhang, H. Liu, and S. Chen, "Real-time Adaptive and Localized Spatiotemporal Clutter Filtering for Ultrasound Small Vessel Imaging," *arXiv preprint arXiv:2405.11105,* 2024.

[61] C. Huang, P. Song, P. Gong, J. D. Trzasko, A. Manduca, and S. Chen, "Debiasing-Based Noise Suppression for Ultrafast Ultrasound Microvessel Imaging," *IEEE Transactions on Ultrasonics, Ferroelectrics, and Frequency Control,* vol. 66, no. 8, pp. 1281-1291, 2019.

[62] H. Wang *et al.*, "Quantitative Assessment of Inflammation in a Porcine Acute Terminal Ileitis Model: US with a Molecularly Targeted Contrast Agent," *Radiology,* vol. 276, no. 3, pp. 809-17, Sep 2015.

[63] W. Schierling *et al.*, "Sonographic real-time imaging of tissue perfusion in a porcine haemorrhagic shock model," *Ultrasound Med Biol,* vol. 45, no. 10, pp. 2797-2804, Oct 2019.

[64] J.-M. Hyvelin *et al.*, "Characteristics and Echogenicity of Clinical Ultrasound Contrast Agents: An In Vitro and In Vivo Comparison Study," *J. Ultrasound Med.,* vol. 36, no. 5, pp. 941-953, 2017.

[65] Z. Zhang *et al.*, "Using 1/2 Descending Time in CEUS to Identify Renal Allograft Rejection," *Acad Radiol,* vol. 31, no. 8, pp. 3248-3256, Aug 2024.

[66] Z. Kou, M. R. Lowerison, Q. You, Y. Wang, P. Song, and M. L. Oelze, "High-Resolution Power Doppler Using Null Subtraction Imaging," *IEEE Trans. Med. Imaging,* vol. 43, no. 9, pp. 3060-3071, 2024.